\documentclass[twocolumn,showpacs,preprintnumbers,prb]{revtex4}

\usepackage{graphicx}
\usepackage{dcolumn}
\usepackage{bm}

\begin{document}

\title{Infrared study of valence transition compound 
YbInCu$_4$ using cleaved surfaces}    

\author{H. Okamura}
\author{T.~Michizawa}
\author{T.~Nanba}
\affiliation{Graduate School of Science and Technology, 
Kobe University, Kobe 657-8501, Japan}

\author{T. Ebihara}
\affiliation{Department of Physics, Faculty of Science, 
Shizuoka University, Japan}

\date{\today}

\begin{abstract}
Optical reflectivity [$R(\omega)$] of YbInCu$_4$ single 
crystals has been measured across its first order valence 
transition at $T_v \simeq$ 42~K, using both polished and 
cleaved surfaces.       
$R(\omega)$ measured on cleaved surfaces [$R_c(\omega)$] 
was found much lower than that on polished surface 
[$R_p(\omega)$] over the entire infrared region.    
Upon cooling through $T_v$, $R_c(\omega)$ showed rapid 
changes over a temperature range of less than 2~K, and 
showed only minor changes with further cooling.  
In contrast, $R_p(\omega)$ showed much more gradual and 
continuous changes across $T_v$, similarly to previously 
reported data on polished surfaces.    
The present result on cleaved surfaces demonstrates that 
the microscopic electronic structures of YbInCu$_4$ 
indeed undergo sudden changes upon the valence transition, 
which is consistent with its first order nature.    
The gradual temperature evolution of $R_p(\omega)$ is most 
likely due to the compositional and/or Yb-In site disorders 
caused by polishing.

\end{abstract}

                             
\maketitle

Physical properties of strongly correlated ``heavy fermion'' 
compounds, most typically Ce- or Yb-based compounds containing 
partly filled 4$f$ shell, have attracted much 
attention.\cite{hewson}      The hybridization between the 
conduction ($c$) electrons and the otherwise localized $f$ 
electrons leads to many interesting phenomena.    
Among them is a duality and crossover between itinerant and 
localized characters of the $f$ electrons.\cite{crossover}   
An important characteristic energy for the $c$-$f$ 
hybridization is the Kondo temperature ($T_K$).    
At temperatures ($T$) well above $T_K$, $f$ electrons are 
basically localized, resulting in local moment (Curie-Weiss) 
paramagnetism.  With decreasing $T$, the $c$-$f$ hybridization 
becomes progressively stronger.   At $T << T_K$, the local 
moment is well screened, and a spatially extended, coherent 
band is formed.    Due to the strong Coulomb correlation of 
$f$ electrons, the resulting bands show enhanced effective 
masses and a Pauli paramagnetism with large effective moment.    
Associated with this crossover from localized to itinerant 
characteristics, the average valence of Ce or Yb ions 
deviates from 3 (decreases for Yb and increases for Ce).

Usually, such a $T$-induced crossover between localized and 
itinerant regimes is continuous, and marked by, e.g., a broad 
maximum in the $T$-dependence of susceptibility and/or the 
$f$ electron-derived resistivity.\cite{crossover}    
However, YbInCu$_4$ undergoes a first-order phase transition 
at $T_v$ $\sim$ 42~K between the high $T$ phase (HTP) in a 
localized regime with $T_K$=20~K and the low $T$ phase (LTP) 
in an itinerant regime with $T_K$=300~K.\cite{felner,sarrao}     
Upon cooling through $T_v$, YbInCu$_4$ shows discrete changes 
in the lattice constant and many other physical properties.     
The Yb valence deduced from measured bulk themodynamic 
properties is almost 3 in HTP, and about 2.9 in LTP.     
It is an intriguing question as to why this particular 
compound shows such a first-order transition between the two 
regimes, while most others show only continuous and gradual 
crossovers.    Several models have been proposed as the 
origin for the transition.\cite{cornelius}

To understand the microscopic nature of such an electronic 
transition, it is important to probe the electronic structures 
around the Fermi level ($E_F$).  For this purpose, infrared 
(IR) reflectivity [$R(\omega)$] of YbInCu$_4$ has been measured 
in detail, at temperatures both above and below 
$T_v$.\cite{marabelli,schle1,schle2,schle3}     
$R(\omega)$ showed large changes between LTP and HTP.    
In the optical conductivity $\sigma(\omega)$ obtained from 
$R(\omega)$, a marked mid-IR peak was observed in LTP, which was 
interpreted\cite{schle1} as arising from optical excitations 
in the $c$-$f$ hybridized state.\cite{c-f}        
This optical result demonstrated that the 
microscopic electronic structures around Fermi level indeed 
undergo large changes upon the valence transition.    
However, the observed $R(\omega)$ seemed to show gradual 
variations upon cooling through $T_v$.\cite{schle3}   
Such a $T$ dependence is in contrast to the first-order 
nature of the transition.

Interesting results have been also reported concerning the 
Yb valence in YbInCu$_4$ studied by photoemission spectroscopy 
(PES).\cite{VUV,SX,HX}     In early PES studies on scraped 
surfaces with a vacuum UV source,\cite{VUV} where the escape 
depth of photoelectrons was small, only a gradual change of 
valence was observed across $T_v$, and the valence was 
much lower than that given by the bulk properties.   
However, as the escape depth increased with soft and hard X-ray 
sources and as the sample was prepaed by cleaving (fracturing), 
a much sharper change of valence was observed, and the valence 
was much closer to the bulk values.\cite{SX,HX}      
These PES results strongly suggest that the valence and its 
transition in YbInCu$_4$ are extremely sensitive to disorder, 
caused either by scraping or by the presence of surface itself.

Motivated by the above developments, we have measured infrared 
$R(\omega)$ of YbInCu$_4$ {\it using both polished and 
cleaved surfaces} of single crystals taken from the same 
batch, and have tried to obtain intrinsic $\sigma(\omega)$ 
of YbInCu$_4$.   
Although the $R(\omega)$ spectra measured on cleaved surfaces 
[$R_c(\omega)$] had qualitatively similar spectral shapes to 
those on polished surfaces [$R_p(\omega)$], the magnitude and 
$T$ dependences were remarkably different between $R_c(\omega)$ 
and $R_p(\omega)$.     
With decreasing $T$, $R_c(\omega)$ showed sudden spectral changes 
at $T_v$, while $R_p(\omega)$ showed only gradual changes.    
The variations of $R_c(\omega)$ are very consistent with the 
first-order nature of the transition, and demonstrate 
that the microscopic electronic structures in YbInCu$_4$ indeed 
undergo a sudden change at $T_v$.      
We analyze the origin of gradual changes in $R_p(\omega)$ 
in terms of microscopic disorder caused by polishing.

The samples used in this work were single crystals grown by 
an In-Cu self-flux method, similarly to that previously 
described.\cite{sarrao}   The resistivity of the single crystal 
decreased rapidly between 41 and 44~K, as shown in Fig.~1(a), 
consistent with the reported value of $T_v$=42~K.    
\begin{figure}[b]
\begin{center}
\includegraphics[width=0.35\textwidth]{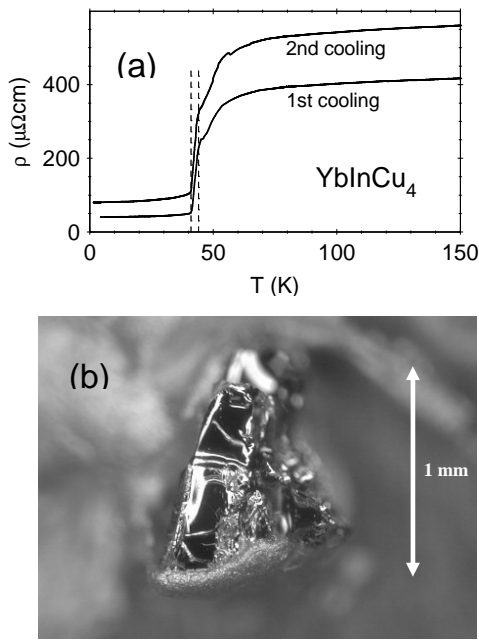} 
\end{center}
\caption{
(a) Resistivity ($\rho$) of YbInCu$_4$ single crystal 
as a function of temperautre ($T$).   Two vertical lines 
indicate 41 and 44~K.  (b) Photograph of cleaved 
(fractured) surfaces of a YbInCu$_4$ single crystal.     
The dark areas indicate flat, specular surfaces.    
}
\end{figure}
The higher resistivity of the second cooling cycle is also 
consistent with that reported previously.\cite{sarrao}     
This increase of resistivity after the first temperature 
cycle was attributed\cite{sarrao} to internal strain caused 
by the lattice contraction upon warming through $T_v$.       
To obtain cleaved surfaces, a block of single crystal was 
fractured into small pieces.     They had flat, specular 
surfaces only with small dimensions (typically 50-500~$\mu$m).   
Those having such specular surfaces were mounted in a 
continuous-flow liquid He cryostat.     The photograph of a 
measured sample is shown in Fig.~1(b).    
$R_c(\omega)$ spectra of thus mounted samples were measured 
using an IR microscope and an IR synchrotron radiation 
(SR) source at the beam line BL43IR of SPring-8.\cite{BL43IR}     
Owing to the high brilliance of IR-SR, it was possible to 
focus the beam to a spot of $\sim$ 15~$\mu$m diameter at the 
sample without using any aperture in the optical path.    
This enabled us to easily measure $R_c(\omega)$ of the small 
specular surfaces of YbInCu$_4$.      
To obtain $R_p(\omega)$, the crystals taken from the same batch 
as those used for $R_c(\omega)$ were mechanically polished.     
$R_p(\omega)$ were measured using an apparatus without a 
microscope, as previously described.\cite{okamura}     
$R_p(\omega)$ were measured up to 30~eV using a SR source at the 
beam line BL7B of UVSOR, Institute for Molecular Science.    
$\sigma(\omega)$ spectra were obtained from the measured 
$R(\omega)$ spectra using the Kramers-Kronig 
relations.\cite{dressel}     
Due to technical restriction, $R_c(\omega)$ spectra were 
measured at 0.06-2 eV only, above which they were smoothly 
connected to $R_p(\omega)$.    Below the measured range, 
$R_c(\omega)$ and $R_p(\omega)$ were extrapolated using 
Hagen-Rubens relation.\cite{dressel}

Figure~2(a) shows $R_c(\omega)$ and $R_p(\omega)$ in HTP and LTP.    
\begin{figure}[b]
\begin{center}
\includegraphics[width=0.38\textwidth]{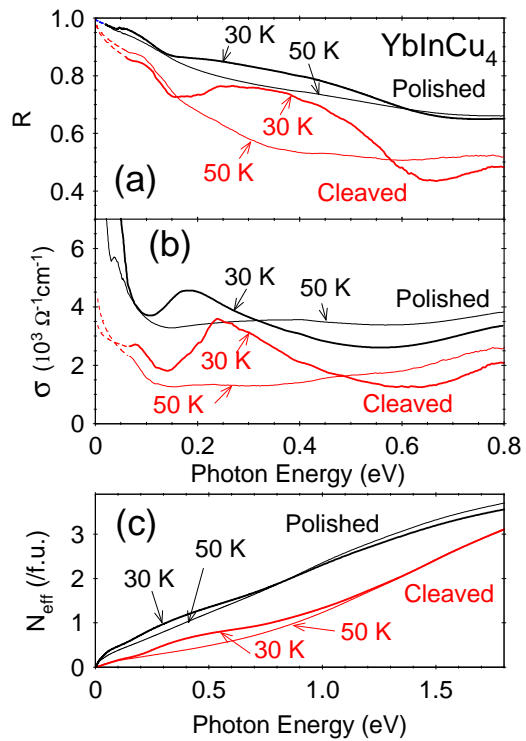}
\end{center}
\caption{(color online)
(a) Reflectivity spectra (R) of YbInCu$_4$ measured on 
cleaved and polished surfaces above and below $T_v$.     
(b) Optical conductivity spectra ($\sigma$) obtained from 
the reflectivity using the Kramers-Kronig relations.  
The spectra in the extrapolated region are indicated by 
broken curves.    
(c) Effective carrier density $N_{\rm eff}$ per formula unit 
(f.u.) calculated from 
$\sigma(\omega)$ using the optical sum rule.    
}
\end{figure}
The spectral shapes of $R_c(\omega)$ and $R_p(\omega)$ are 
qualitatively similar to each other.    Namely, in the 
high-$T$ phase the spectrum is a decreasing function 
of photon energy.    In the low-$T$ phase, on the other hand, 
the reflectivity at 0.2-0.55~eV increases significantly and shows 
a marked dip centered at $\sim$ 0.15~eV.      These spectral 
variations are qualitatively similar to previous 
results.\cite{schle1,schle2,schle3}     
However, the overall magnitude of $R_c(\omega)$ is much lower 
than that of $R_p(\omega)$.    
We have carefully confirmed that the low value of $R_c(\omega)$ 
is not due to experimental errors, but an intrinsic property 
of the cleaved surfaces.\cite{footnote3}    In addition, 
the difference in $R_c(\omega)$ between the two phases is 
much larger than that in $R_p(\omega)$.   
Figure~2(b) shows the optical conductivity spectra of the 
cleaved [$\sigma_c(\omega)$] and polished [$\sigma_p(\omega)$] 
samples, obtained from $R_c(\omega)$ and $R_p(\omega)$, 
respectively.     
They show large difference in their magnitude, correspoonding 
to that between $R_c(\omega)$ and $R_p(\omega)$, while their 
spectral shapes appear qualitatively similar to each other.   
It appears as though $\sigma_p(\omega)$ results from the 
superposition of a constant background upon $\sigma_c(\omega)$.    
Note, however, that the mid-IR peak energy of 0.25~eV in 
$\sigma_c(\omega)$ is about 70~meV higher than in 
$\sigma_p(\omega)$.     In previous works on polished samples, 
the mid-IR peak was observed at $\sim$ 0.25~eV,\cite{schle2,schle3} 
and at $\sim$ 0.3~eV ($\sim$ 2500~cm$^{-1}$).\cite{schle1}    
Figure~2(c) shows the effective carrier density 
$N_{\rm eff}(\omega) = (2 m_0/ \pi e^2) \int_{0}^{\omega} 
\sigma(\omega^\prime)d\omega^\prime$ given by the optical 
sum rule.\cite{dressel}     
The sum rule is satisfied (the spectral weight transfer is 
completed) at 1.3~eV for the cleaved sample, since 
$N_{\rm eff}$ below $T_v$ is almost equal to that 
above $T_v$ for $\hbar \omega \geq$ 1.3~eV.   In contrast, 
it is not satisfied up to 1.8~eV for the polished sample.

Figure~3 compares the detailed $T$ dependences of $R_c(\omega)$ and 
$R_p(\omega)$.   
\begin{figure}[b]
\begin{center}
\includegraphics[width=0.38\textwidth]{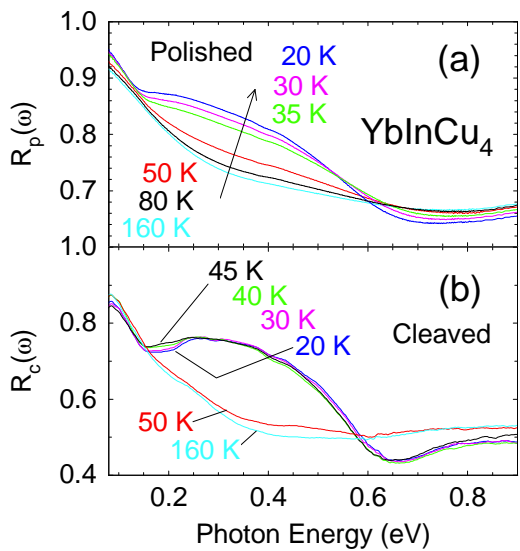}
\end{center}
\caption{(color online)
Reflectivity spectra of YbInCu$_4$ single crystals 
measured on (a) polished and (b) cleaved surfaces.    
Note that the vertical scale is different between (a) and (b).   
}
\end{figure}
$R_p(\omega)$ changes gradually with decreasing $T$, 
and keeps changing even below 40~K.      
These spectral evolutions agree well with those previously 
reported for polished surfaces.\cite{schle3}    
The $R_c(\omega)$ spectra, 
on the other hand, show large changes over a very narrow 
$T$ range.     In addition, $R_c(w)$ is almost unchanged 
in LTP, with only minor variations in the shape of the dip.     
Clearly, the $T$-dependence of $R_c(\omega)$ is quite 
different from that of $R_p(\omega)$.      In Fig.~4, 
to show their $T$ variations more quantitatively, 
$R_c(\omega)$ and $R_p(\omega)$ integrated at 0.2-0.55~eV 
are plotted as a function of $T$.     
\begin{figure}[t]
\begin{center}
\includegraphics[width=0.40\textwidth]{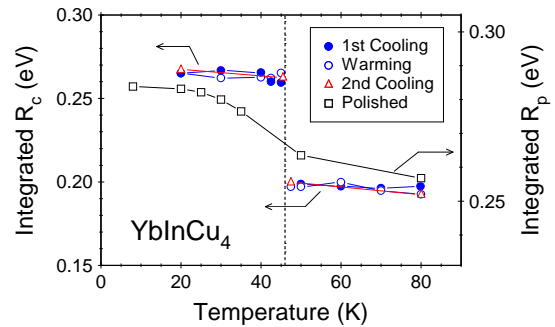}
\end{center}
\caption{(color online) 
$R_c(\omega)$ and $R_p(\omega)$ integrated from 0.2 
to 0.55~eV as a function of temperature.     The vertical 
line is drawn at 46~K, and is a guide to the eye.    
}
\end{figure}
It is seen that the change in $R_c(\omega)$ occurs at $\sim$ 46~K 
over a $T$ range of less than 2~K, in contrast to the gradual 
change of $R_c(\omega)$.\cite{footnote2}    
While the $T$ variation in $R_c(\omega)$ is as sharp as that 
in the resistivity, the transition $T$ of 46~K implied 
by $R_c(\omega)$ is slightly higher than 41-44~K implied by 
the resistivity [Fig.~1(a)].   
The variations of integrated $R_c(\omega)$ 
with $T$ was accompanied by a hysterisis of about 1~K.  
Such a hysteresis is consistent with those 
observed in the thermodynamic properties.     
Another interesting feature in Fig.~4 is that no significant 
diffrence is observed in $R_c(\omega)$ between the first and 
second cooling cycles, in contrast to the different 
resistivities for the first and second coolings 
[See Fig.~1(a)].

Clearly, the $T$ evolution of $R_c(\omega)$ is more consistent 
with the first-order naure of the valence transition, and with 
the $T$ variations of other physical properties such as 
resistivity and susceptibility, which change suddenly upon 
the transition and remain almost constant below $T_v$.     
Hence the $R_c(\omega)$ data should reflect the intrinsic bulk 
electronic structures of YbInCu$_4$ more closely and 
directly than the $R_p(\omega)$ data.     However, as mentioned 
before, the spectral shape of $\sigma_c(\omega)$, in particular 
the mid-IR peak in LTP, is qualitatively similar to that in 
$\sigma_p(\omega)$.    
Hence the interpretation of $\sigma(\omega)$ in LTP in terms 
of a $c$-$f$ hybridized state reported in the previous 
works\cite{schle1,schle2,schle3} should remain valid.     
The origin for the spread of mid-IR peak energies 
(0.18-0.3~eV) for polished samples in this work 
and in the previous works\cite{schle1,schle2} is unclear.    
It may have resulted from differenes in the specific way 
how the polishing was done in these works.     
On the other hand, the magnitude and $T$-dependence of 
$\sigma_c(\omega)$ are quite different from those of 
$\sigma_p(\omega)$.       Therefore the previous data 
on the spectral weight and its $T$-induced transfer in 
$\sigma(\omega)$ measured for polished YbInCu$_4$ should 
be interpreted carefully.\cite{schle1,schle3}

The broadened $T$ dependence in $R_p(\omega)$ should be due to 
disorder introduced by the mechanical polishing.     
Two kinds of disorder are likely to be responsible: a deviation 
from the ideal 114 composition and a site disorder.      
Since the early stage of research on YbInCu$_4$, it has been 
recognized that the $T_v$ value for Yb$_{1-x}$In$_{1+x}$Cu$_4$ 
is a sensitive function of $x$.\cite{felner}     
Such a compositional disorder is quite likely to result 
from mechanical polishing, which may lead to a 
distribution of $T_v$ within the 
penetration depth of the IR radiation.    
The penetration depth of YbInCu$_4$ in the photon energy 
range of 0.1-0.2~eV has been estimated to be 
500-1000~{\AA} from the absorption coefficient obtained 
with the Kramers-Kronig analysis of 
$R_c(\omega)$ and $R_p(\omega)$.   
In addition, even in high-quality YbInCu$_4$ samples with 
a negligible compositional deviation, a site disorder tends 
to occur between the In and Yb sites (Yb occupying In site, 
and vice versa).\cite{disorder}     
It has been pointed out that such a site disorder may change 
the Yb valence, and the $T_v$ value.\cite{disorder}    
It is likely that the surface layer of a mechanically polished 
sample has more site disorder, and hence a broader transition, 
than that of a cleaved surface. 
We believe that the broadened $T$ variation of $R_p(\omega)$ 
results from the compositional and/or site disorder.

Note that the sudden change of $R_c(\omega)$ 
occurs at 46~K, in contrast to 41-44~K in the resistivity.   
This is a rather unexpected result, since it is natural to 
expect $R(\omega)$ and resistivity to change at the 
same transition temperature.    
The origin for this diffrence is unclear at the present time.    
One possibility is that $T_v$ is sensitive even to the cleaving, 
so that the $T_v$ becomes slightly higher at the near-surface 
region [without, however, a distribution of $T_v$ since the 
$T$ variation of $R_c(\omega)$ itself is very sharp].   
Further study is needed to clarify this point.

In addition to the different $T$ dependences, $R_c(\omega)$ has 
much lower magnitude than $R_p(\omega)$ at all measured $T$'s, 
as mentioned above.        
Since the higher $R_p(\omega)$ has been caused by polishing, 
it is also likely due to disorder.     
The higher $R_p(\omega)$ probably results from higher 
reflectivity of composition-disordred or site-disordered 
portion of YbInCu$_4$ in the surface layer.   
It is even likely that some fraction of the disordered 
portion at surface does not undergo the valence transition, 
hence contributing a $T$-independent, constant backgraound 
in $R(\omega)$ and $\sigma(\omega)$.

In conculsion, IR reflectivity spectra of YbInCu$_4$ 
have been measured on cleaved and polished surfaces of 
single crystals.   The reflectivity spectrum measured 
on cleaved surface showed sudden changes over a narrow 
$T$ range upon the valence transition, in contrast to the 
gradual $T$ evolution observed for polished surfaces.    
The result on cleaved surface demonstrates that the 
microscopic electronic structures in YbInCu$_4$ indeed 
undergo a first-order transition.   
The broadened $T$ evolution of reflectivity for polished 
sample has been analyzed in terms of compositional and/or 
Yb-In site disorder caused by polishing.

H. O. thanks Jason Hancock for useful communications and 
the staff at SPring-8 for technical assistance.   
Experiments at UVSOR were done as a Joint Studies Program of 
IMS, and those at SPring-8 under the approval by JASRI 
(2004A0778-NSa-np).    This work has been supported by 
Grants-in-Aid for Scientific Research No.~17340096 and 
Priority Area ``High Field Spin Science in 100~T'' 
(No.~451) from  MEXT.    



\begin{thebibliography}{99}
\bibitem{footnote2}
Although the $R_p(\omega)$ spectra were recorded only at 
selected $T$'s, the total reflection intensity was 
continuously monitored throughout the cooling and warming 
processes.    No steep change of the reflectivity 
was observed around $T_v$, unlike the case of $R_c(\omega)$.
\end{thebibliography}
\end{document}